\documentstyle{mn}

\def\gta{\ifmmode {\mathbin{\lower 3pt\hbox   
    {$\,\rlap{\raise 5pt\hbox{$\char'076$}}\mathchar"7218\,$}}}
    \else {${\mathbin{\lower 3pt\hbox
    {$\rlap{\raise 5pt\hbox{$\char'076$}}\mathchar"7218\,$}}}
    $}\fi}
\def\lta{\ifmmode {\,\mathbin{\lower 3pt\hbox   
    {$\,\rlap{\raise 5pt\hbox{$\char'074$}}\mathchar"7218\,$}}}
    \else {${\mathbin{\lower 3pt\hbox
    {$\rlap{\raise 5pt\hbox{$\char'074$}}\mathchar"7218\,$}}}
    $}\fi}

\title{Production of Intermediate-Mass Black Holes in Globular Clusters}
\author[M. Coleman Miller and Douglas P. Hamilton]
       {M. Coleman Miller and Douglas P. Hamilton\\
        Department of Astronomy,
        University of Maryland, College Park, MD 20742-2421}

\begin{document}

\maketitle

\label{firstpage}

\begin{abstract}

The discovery of numerous non-nuclear X-ray point sources with
luminosities $L>10^{39}\,{\rm erg~s}^{-1}$ in several starburst galaxies
has stimulated speculation about their nature and origin.  The strong
variability seen in several sources points to massive black holes as the
central engines.  If the flux is isotropic, the luminosities range up to
$\approx 10^{41}$~erg~s$^{-1}$, implying masses of $M\gta 10^3\,M_\odot$ if
the luminosity is sub-Eddington. Here we explore a model for these
sources.  We suggest that in some tens of percent of globular clusters a
very massive black hole, $M\gta 50\,M_\odot$, is formed.  This black hole
sinks in $\lta 10^6$~yr to the center of the cluster, where in the $\sim
10^{10}$~yr lifetime of the cluster it accretes $\sim 10^3\,M_\odot$,
primarily in the form of lighter black holes.  Unlike
less massive black holes in binaries, which are flung from clusters by recoil
before they can merge gravitationally, a $\gta 50\,M_\odot$ black hole has
enough inertia that it remains bound to the cluster.  We suggest that
$\sim 10^3\,M_\odot$ black holes may be common in the centers of
dense globular clusters, and may therefore exist in some tens of percent
of current globulars.  If the cluster later merges with its host galaxy,
accretion from young star clusters in molecular clouds by the black hole
can generate luminosity consistent with that observed. We also consider
the detectability of massive black holes in globular clusters with
gravitational wave detectors such as LISA and LIGO, and speculate on
future observations that may test our predictions.

\end{abstract}


\section{Introduction}

In the last few years evidence has mounted for a  population of spatially
unresolved sources in starburst galaxies that have fluxes corresponding
to isotropic X-ray
luminosities in excess of $\sim 10^{39}$~erg~s$^{-1}$ (e.g., Fabbiano
1989; Fabbiano, Schweizer, \& Mackie 1997; Colbert \& Mushotzky 1999;
Zezas, Georgantopoulos, \& Ward 1999; Kaaret et al. 2001; Matsumoto et
al. 2001; Fabbiano, Zezas, \& Murray 2001).  The variability of these
sources on time scales as short as $10^4$~s (Matsumoto et al. 2001)
suggests that they are accreting compact objects, probably black holes.
If beaming of emission is moderate to negligible, the brightest of these
objects (with $L\approx 10^{41}$~erg~s$^{-1}$) must have masses
$M\gta 10^3\,M_\odot$ so that the luminosity is below the Eddington
luminosity $L_E=1.3\times 10^{38}\,(M/1\,M_\odot)$~erg~s$^{-1}$. Their
non-nuclear locations imply an upper mass limit as low as
$10^{5-6}\,M_\odot$, otherwise dynamical friction would have caused them
to sink to the center of their host galaxies (Kaaret et al. 2001). The
inferred masses are much larger than the $\sim 10-20\,M_\odot$ masses of
black holes usually thought to arise from the evolution of present-day
stars (Fryer 1999), yet are much smaller than the  $\sim
10^{6-9}\,M_\odot$ masses of black holes that exist in the centers of
many galaxies (e.g., Merritt \& Ferrarese 2001).   The origin of such
objects is an intriguing mystery.  It has been suggested that the
emission is beamed (King et al. 2001), so that the accreting black holes
have ordinary stellar masses, or that these black holes formed directly
via evolution of Population III stars (Madau \& Rees 2001).

Here we propose another scenario, in which a black hole grows
significantly by accretion.  We show that for black holes with
initial mass $M\lta 10^3\,M_\odot$, accretion from the ISM is too slow
to increase the mass significantly. Accretion from stars in a dense
cluster is more promising (see also Taniguchi et al. 2000), but the
cluster must last several billion years for significant growth, which
points to globular clusters.  However, a $\sim 10\,M_\odot$ black hole
sinks to the cluster core within $\lta 10^7$~yr, where it forms a binary
or exchanges into one via three-body processes.  Further three-body
interactions tighten the binary, but each tightening causes the binary
to recoil. Before the binary becomes hard enough to merge rapidly via
gravitational radiation, the recoil kicks it completely out of the
cluster (Sigurdsson \& Hernquist 1993).  Thus, a light black hole
usually cannot increase its mass by a large factor despite being in a
high-density stellar environment.

In contrast, as we show in \S~2, a $\gta 50\,M_\odot$ black hole is so
massive that recoil does not eject it from the cluster before the binary
merges.  Hence, the process can be repeated and in a dense cluster a
$\gta 50\,M_\odot$ black hole can increase its mass to $\sim
10^3\,M_\odot$.  This has some similarities to one of the ways in
which supermassive black holes may grow in the nuclei of galaxies (see,
e.g., Frank \& Rees 1976; Sigurdsson \& Rees 1997).

We suggest that $\sim 10^3\,M_\odot$ black holes are generated in
this fashion in some tens of percent of clusters. While in the clusters,
accretion onto the black hole is usually electromagnetically quiet, since
most accretion is of compact objects that are swallowed whole. However,
if the cluster merges with the disk of the host galaxy, as may occur for
roughly half of globulars (e.g., Gnedin \& Ostriker 1997; Gnedin, Lee, \&
Ostriker 1999; Takahashi \& Portegies Zwart 1998, 2000), 
then after the globular has settled into the mean motion
of the disk and has been disrupted by galactic tidal fields, the massive
black hole is released.  If it later enters a molecular cloud, the extra
mass of the hole may speed up gravitational contraction and star
formation, resulting in the black hole being at the center of (and
accreting from) a cluster of stars.  For starburst galaxies, where the
number of molecular clouds and volume-averaged density of interstellar
gas may be a few times the corresponding values in normal galaxies,
several such sources can be detectable at any given time.

In \S~2 we  estimate the growth rate of such black holes and the duty
cycle of high-luminosity accretion (and hence detectability in globular
clusters).  In \S~3 we calculate the expected accretion rate once in the
disk of the host galaxy.  In \S~4 we discuss the implications of this
model for the upcoming generation of gravitational wave detectors, and
we present our conclusions and future observational tests in \S~5.

\section{Formation and Growth of Black Holes in Globular Clusters}

\subsection{Birth of the initial $\gta 50\,M_\odot$ black hole.}

Substantial growth of a black hole by mergers in a globular cluster
requires that it have a high enough mass, $M\gta 50\,M_\odot$, to avoid
being kicked out by recoil during three-body interactions (see
\S~2.2).  How can such
black holes form?  We identify three possible mechanisms.  The first is
direct formation from a massive star.  A number of recent numerical studies
indicate that if a star has a mass $M\gta 30-40\,M_\odot$ when its core
collapses then the resulting explosion may not have enough energy to drive
off the outer layers of the star (e.g., Fryer 1999).  If so, and if the
star has low enough angular momentum, then direct fallback from the star
may produce a black hole with a mass close to that of the original star
(Fryer 1999).  For a star with an initial mass of $\sim 50-100\,M_\odot$,
the question is then whether stellar winds and mass ejection during the
main sequence and giant phases  are enough to reduce the mass well below
the $\sim 50\,M_\odot$ threshold.  The amount of mass lost in a wind is
challenging to  determine, but some recent calculations suggest that,
especially for stars in a low-metallicity environment such as a globular
cluster, the wind loss is minor; for example, from Table 3 of Vink, de
Koter, \& Lamers (2001), the mass loss rate for a $40\,M_\odot$ star with
metallicity $Z={1\over{30}}Z_\odot$ is less than
$10^{-7}\,M_\odot$~yr$^{-1}$,  implying a total of less than $1\,M_\odot$
during the lifetime of the star.  The number of such stars is difficult to
estimate, given the poor statistics.  Observations of stars in young
clusters imply that for $M>1\,M_\odot$ the Salpeter mass function
$dN/dM\propto M^{-2.35}$ is a decent approximation, with no hint of a
breakdown for stars even as massive as $\sim 100\,M_\odot$ (Massey 1998).
This would imply $\sim 10^{-4}$ of stars have $M>50\,M_\odot$, and hence
that a globular cluster will potentially have tens of this mass.

The second candidate mechanism is a variant of the first.  In a dense
stellar environment the most massive stars sink to the center of a
cluster.  If this happens more rapidly than the stars evolve off the main
sequence (e.g., a few million years for a $30\,M_\odot$ star), the stars may then
undergo frequent collisions and mergers (Portegies Zwart et al. 1999),
which produce more massive stars.  Recent analyses (Fryer 1999) suggest
that when these massive stars undergo core collapse, they may leave
$50-100\,M_\odot$ black holes, even when mass loss from strong stellar
winds is included.

The third possibility is that if hundreds to thousands of $\sim 10\,M_\odot$
black holes are formed in a cluster, there will be enough interactions
that a few will merge with each other, forming a (smaller) set of
$\sim 20\,M_\odot$ black holes, and so on, until the $\sim 50\,M_\odot$
threshold is reached.  In this scenario, the overwhelming majority
of small black holes are ejected from the cluster; it is only the
fortunate minority that can merge.  This mechanism has apparently not
been explored in detail.

\subsection{Growth of $\gta 50\,M_\odot$ black holes in cluster cores.}

Black holes formed from supernovae are much more massive than the
average mass $\langle m\rangle\approx 0.4-0.6\,M_\odot$ of stars in
globular clusters, hence they sink rapidly to the center of the
cluster.  A black hole of mass $M$ sinks to the center in a time
$\sim \langle m\rangle/M$ times the core half-mass relaxation time $t_r$.
For typical globular clusters, $t_r\sim 10^{7-9}$~yr.  Thus, 
a $\gta 50\,M_\odot$ black hole will
sink to the center within $\lta 10^6$~yr, shorter than the
other time scales of interest.

The rate of encounters in the core obviously depends on the number density
of field black holes in the vicinity.  Numerical simulations of globular
clusters (Sigurdsson \& Phinney 1995) show that in the cores of clusters in
thermal equilibrium  with number densities $n\gta 10^5$~pc$^{-3}$ encounters
of binaries with single stars are dominated by the most massive field stars,
in this case black holes.  This is because in thermal equilibrium the
massive stars have a low velocity, hence they sink to the center of the
cluster and have a high number density even if their total numbers are
comparatively small.  For example, in a multimass King model the scale
height of a stellar species of mass $m$ scales as $(\langle m\rangle/m)^{1/2}$
(see, e.g., Sigurdsson \& Phinney 1995).  Thus, black holes with typical
mass $10\,M_\odot/0.4\,M_\odot=25$ times the average mass of a main sequence
star have a scale height 1/5 of the scale height of the typical star, and
hence a number density enhanced by a factor of $\sim 100$.  Therefore, if
their number fraction averaged over the cluster is more than $\sim 1$\%, their
core number density is comparable to or greater than that of visible stars.
In addition, black holes may settle gravitationally with enough
rapidity that they are no longer in thermal equilibrium with the lower-mass
stars (e.g., Meylan 2000); in this case, their number density in the core
and hence all encounter rates are correspondingly increased.  For simplicity
we assume that the core number density of black holes equals that of
main sequence stars, but there is considerable uncertainty in this number.

We focus on clusters that are at or near core collapse, and adopt
$10^6n_6$~pc$^{-3}\approx 3\times 10^{-50}n_6$~cm$^{-3}$ as a fiducial
number density of normal black holes near the massive black hole.  Of the 56
Galactic globular clusters studied by Pryor \& Meylan 1993 (out of a total
of 143;  see Djorgovski \& Meylan 1993), 6 had central mass densities
greater than $10^6\,M_\odot$~pc$^{-3}$ and 22 had central mass densities
greater than $10^5\,M_\odot$~pc$^{-3}$.  The average stellar mass is less
than $1\,M_\odot$ (Pryor \& Meylan 1993), so the number densities are
correspondingly higher.  We show in \S~2.2.2 that if $n_6\gta 0.1$ the cores
of clusters are favourable for the growth of black holes to masses $\sim
10^3\,M_\odot$, so this implies that some 40\% of Galactic globulars, or
$\approx 60$, could harbour such black holes.  Note that this could be
a substantial underestimate; for our purposes the important quantities
are the masses and densities of clusters at birth, but to be conservative
we are guided by the current properties of Galactic globulars.  As
mentioned by Portegies Zwart \& McMillan (2000), zero age globulars may
have had $\sim$3 times the mass and $\sim 1/3$ of the virial radius of
current globulars, and hence processes leading to the formation of massive 
black holes are likely to have been significantly more efficient in the past.

While in the high stellar density environment of the core, the black hole
undergoes a variety of interactions that increase its mass via merger or
accretion.  If there is a sufficient fraction of binaries in the core
(quantified in \S~2.2.1), the massive black hole exchanges into a binary and
ultimately merges.  If the number of binaries is low, the black hole can
capture a field star through tidal interactions (for a main sequence
star) or by gravitational radiation (for a compact object).  Three-body
interactions strongly favour the resulting
binary being composed of the two most massive stars of the three that
interact.  The most massive field stars could be $\sim 10\,M_\odot$ main
sequence stars within the  first $\sim 10^8$~yr of the cluster or, later
on, could be $\sim 10\,M_\odot$ black holes (if these are abundant),
neutron stars, or white dwarfs. Further interactions tend to harden the
binary, but they also tend to eject the third, field star.   If instead
accretion is dominated by captures of field compact objects by emission
of gravitational waves, the compact objects are accreted without ejection.  
We now quantify these scenarios,
starting with the demarcation between binary rich and binary poor.

\subsubsection{Division between binary-rich and binary-poor cluster cores}

Which occurs more often, direct capture or binary exchange? A hard binary
is defined to have an orbital binding energy greater than the kinetic
energy of a typical field star.  Assuming a velocity dispersion of
$10^6v_{\rm ms,6}$~cm~s$^{-1}$ for $0.4\,M_\odot$ field stars, this
implies an orbital separation of $\sim 10^{15}$~cm for black hole binaries
of total mass $\sim 10\,M_\odot$.  Thus, if a massive black hole
encounters such a binary with a periastron $r_p$ less than $10^{15}$~cm,
there is likely to be a strong interaction with an exchange.  This is
$\approx 10^4$ times the $\sim 10^{11}$~cm periastron necessary for direct
capture (see \S~2.2.3); since in the strong focusing limit the cross
section scales linearly with the periastron, $\sigma_{\rm coll}\approx \pi
r_p (2GM/v_\infty^2)$ (where $v_\infty$ is the velocity of the field star
at infinity), this implies that if the binary fraction is much larger than
$\sim 10^{-4}$ in the core, a $\sim 50\,M_\odot$ black hole will interact
with binaries far more frequently than it will capture single stars.

\subsubsection{Binary-rich clusters}

Now consider the case in which core binaries are plentiful by the above
definition.  This may be true even if the primordial binary fraction is low,
due to processes such as three-body binary formation (e.g., Lee 1995).  If a
hard binary is formed,  subsequent close three-body interactions pair the
massive black hole with the highest-mass species present in abundance (a
consequence of exchange interactions), and tend to harden the binary further
(Heggie 1975).  However, for a stellar black hole of ``normal" mass $M\lta
10\,M_\odot$, the recoil of the binary that accompanies its hardening ejects
the binary from the cluster before the binary separation becomes small
enough that merger by gravitational radiation occurs faster than the next
binary encounter (Sigurdsson \& Hernquist 1993).  This may not be true for a
more massive black hole.

{\it Ejection or retention of binaries.}---To estimate the threshold mass,
we adopt the treatment of Portegies Zwart \& McMillan (2000).  We are
interested here in binaries with very unequal mass, but we assume for
simplicity that the field black hole has a mass comparable to the mass of
the lighter hole in the binary.  For three equal-mass stars the binding
energy is usually increased by $\sim 20$\% by a strong interaction (e.g.,
Heggie 1975; Sigurdsson \& Phinney 1993), but for high mass ratio binaries
Quinlan (1996) finds that in a given hardening interaction the binding
energy of the binary typically increases by a fraction $\sim
0.2(m/M)=0.04m_{10}M_{50}^{-1}$, where the mass of the large black hole is
$50M_{50}\,M_\odot$ and of the smaller black holes is $10m_{10}\,M_\odot$.
The fraction of the change in binding energy carried away by binary recoil
also depends on the mass ratio.  For three equal-mass objects, conservation
of momentum implies that the kinetic energy of recoil of the binary is
approximately a third of the change in binding energy. If the binary is
much more massive than the third star, $M\gg m$, then the kinetic energy of
the binary is instead roughly $m/M$ times the change in binding energy.
Let $v_{\rm esc}$ be the escape velocity from the core, where typically
$v_{\rm esc}\approx 50$~km~s$^{-1}$ for a dense cluster (Webbink 1985).
Then the binary binding energy that typically produces recoil at the
escape velocity is
\begin{equation}
E_{\rm b,min}\approx 5(M/m)^2\left({1\over 2}Mv_{\rm esc}^2\right)\approx
2\times 10^{50}M_{50}^3m_{10}^{-2}\,{\rm erg}\; .
\end{equation}
This needs to be compared with the binding energy such that gravitational
radiation causes a merger before the next three-body interaction.  The
merger time for two stars with total mass $M$ and reduced mass $\mu$ in an
orbit of semimajor axis $a$ and eccentricity $e$ is (Peter 1964)
\begin{equation}
\tau_{\rm merge}=3\times 10^8M_\odot^3(\mu M^2)^{-1}(a/R_\odot)^4
(1-e^2)^{7/2}\,{\rm yr}.
\label{inspiral}
\end{equation}

From numerical simulations of three equal-mass black holes (Portegies
Zwart \& McMillan 2000), the typical eccentricity distribution  of
binaries after the interaction of three equal-mass objects 
is $P(e)=2e$ (with a slight excess
at high eccentricities).  With this distribution, the average eccentricity
weighted over merger time is $e\approx 0.7$. We note, however, that the
strong dependence of merger time on eccentricity means that deviations
from this distribution can have large effects on the rate of growth of
large black holes in globular clusters, the detectability of gravitational
radiation from these systems, and other important quantities.  Quinlan
(1996) finds that when perturbations are small (as they are when the mass
ratio is high) the eccentricity of a hard binary is increased steadily by
the hardening process until gravitational radiation becomes important.  If
so, merger times are decreased and conditions are more favourable for the
growth of $\sim 10^3\,M_\odot$ black holes in the centers of globular
clusters.  However, in the presence of finite perturbations the outcome is
not as clear.  An important future project is therefore a numerical study
of the growth of $\gta 50\,M_\odot$ black holes in globulars, with a full
treatment of the evolution of the eccentricity.  In the meantime, we
assume conservatively that $P(e)=2e$ regardless of the amount of hardening
that has occurred, until the point that gravitational radiation is
significant.

The typical semimajor axis for a merger time
$\tau_{\rm merge}= 10^6\tau_6$~yr and $M\gg m$ is then
\begin{equation}
a\approx 3\times 10^{11}\tau_6^{1/4}M_{50}^{1/2}m_{10}^{1/4}~{\rm cm}\; .
\end{equation}
The binding energy is
\begin{equation}
E_{\rm bind,merge}=
GMm/a\approx 4\times 10^{50}\tau_6^{-1/4}M_{50}^{1/2}m_{10}^{3/4}\,{\rm erg}\;.
\end{equation}
Therefore, the ratio of
the binding energy necessary to eject the binary from the cluster
to the binding energy necessary for a fast merger is
\begin{equation}
x=E_{\rm b,min}/E_{\rm bind,merge}\approx 0.5\,M_{50}^{5/2}m_{10}^{-11/4}
\tau_6^{1/4}\; .
\end{equation}
If $x>1$, mergers happen before ejection, so that the hole gains
mass and remains near the mass source.  If $x<1$, ejection happens
first, so that although the binary might merge in less than a Hubble
time it will do so well away from the high stellar density in the
cluster, meaning that the black hole mass growth is stopped.  This
calculation shows that if $M\gta 50\,M_\odot$ the hole will stay
near the center of the potential and grow rapidly, whereas for much
smaller masses the hole will be ejected quickly without additional
mass growth.  If the eccentricity increases as the orbit shrinks, the
critical semimajor axis increases, and hence the ratio
$x$ increases.  Therefore, eccentricity growth allows lower-mass black
holes to grow by mergers without being ejected.

{\it Time scale for growth by three-body interactions.}---The 
interaction cross section for a massive black hole is dominated
by gravitational focusing, so that
$\sigma_{\rm coll}\approx \pi r_p(2GM/v_\infty^2)$.
The relevant pericenter distance depends on the type of
encounter and the nature of the secondary object.  For example,
a $10\,M_\odot$
BH companion with a typical eccentricity $e=0.7$ and a semimajor axis
$\sim 5\times 10^{11}\tau_6^{1/4}$~cm will
merge via gravitational radiation within $10^6\tau_6$~yr.
A third black hole passing
within $\sim$twice that distance will produce a strong interaction,
so $r_p\sim 10^{12}$~cm.  In thermal equilibrium the velocity dispersion
of stars is inversely proportional to the square root of their mass, so
if the average mass of a main sequence star is $0.4\,M_\odot$ and its
velocity is $v_{\rm ms}=10^6v_{\rm ms,6}$~cm~s$^{-1}$ then black holes
have a velocity dispersion $v\approx {1\over 5}m_{10}^{-1/2}v_{\rm ms}=
2\times 10^5m_{10}^{-1/2}v_{\rm ms,6}$~cm~s$^{-1}$.
Then $\sigma_{\rm coll}\approx
10^{30}(r_p/10^{12}\,{\rm cm})M_{50}m_{10}$~cm$^2$, so
the time between interactions is
$\tau_{\rm int}=1/\langle n\sigma v_\infty\rangle\approx 2\times 10^6 v_{ms,6}
n_6^{-1}r^{-1}_{p,12}M_{50}^{-1}m_{10}^{-1/2}$~yr,
where $r_p=10^{12}r_{p,12}$~cm and the average in angle brackets is
taken over the velocity distribution of black holes (assumed to be
Maxwellian).  The binding energy scales with $1/a$, as does the
time between interactions.  Therefore, the orbit starts shrinking
rapidly, but slows down.  The total time required to reach $a=
5\times 10^{11}$~cm
from an initially much larger semimajor axis is $\sim 5(M/m)\tau_{\rm int}$
if a fraction $\sim 0.2(m/M)$ of the binding energy is removed in each
close encounter.  The orbital separation for a fixed merger time
scales as $M_{50}^{1/2}$, so $\tau_{\rm int}\sim M_{50}^{-3/2}$ and a
$\sim 50\,M_\odot$ black hole will increase its mass by $\sim 10
\,M_\odot$ in a total time 
$\sim 5\times 10^7n_6^{-1}M_{50}^{-1/2}m_{10}^{-5/2}$~yr.
Integrating, after time $t$ the mass of the black hole is
$M_{50}\approx \left[t/\left(3\times 10^8\,n_6^{-1}m_{10}^{-5/2}\,{\rm yr}
\right)\right]^2$.  Within a Hubble time there is therefore
time for substantial growth in the mass of large black holes in
the centers of clusters that are at least moderately dense, with
$n_6\gta 0.1$.  If, as argued by Quinlan (1996), three-body effects
increase the eccentricity as the binary hardens, the
rate of increase of black hole mass is enhanced significantly.

\subsubsection{Binary-poor clusters}

Now consider what happens when binary 
interactions and exchanges can be neglected.
The effective capture cross section of a compact object of
mass $m$ by a large black hole of mass $M\gg m$ is (Quinlan \& Shapiro 1989)
\begin{equation}
\sigma= 2\pi\left(85\pi\over{6\sqrt{2}}\right)^{2/7}
{G^2m^{2/7}M^{12/7}\over{c^{10/7}v_\infty^{18/7}}}\; .
\end{equation}
Here $v_\infty$ is the relative velocity at infinity and $c$ is the speed
of light.  Numerically, $\sigma\approx 2\times 10^{26}
m_{10}^{2/7}M_{50}^{12/7}v_6^{-18/7}\,{\rm cm}^2$, where
$v_\infty=10^6v_6$~cm~s$^{-1}$.  The strong dependence on
relative velocity means that for equal number densities the slowest
moving population will dominate the encounter rate.  In thermal 
equilibrium, this means the black holes.  Averaging the encounter rate 
$\langle n\sigma v_\infty\rangle$  over the velocity distribution
(assumed Maxwellian), this implies  $\sigma\approx 2\times
10^{28}m_{10}^{11/7}M_{50}^{12/7} v_{\rm ms,6}^{-18/7}\,{\rm cm}^2$
(recall that $v_\infty\propto m^{-1/2}v_{\rm ms}$).   The periastron is
only $\sim 10^{11}$~cm for such an encounter, implying a merger time of
$\sim 10^4$~yr once the orbit is circularised (see
equation~(\ref{inspiral});  circularisation also takes place rapidly).
Thus, if a compact object is captured in this manner it is swallowed
almost immediately, without further interactions.

  The encounter rate for a massive black hole is then 
\begin{equation}
\nu_{\rm enc}=\langle n\sigma v_\infty
\rangle\approx 2\times 10^{-16}
n_6m_{10}^{11/7} M_{50}^{12/7}v_{\rm ms,6}^{-11/7}\,{\rm s}^{-1}\; .
\end{equation}
The characteristic time scale to increase the black hole's mass by
of order itself is then $t_{\rm acc}\approx (M/m)\nu_{\rm enc}^{-1}$, or
\begin{equation}
t_{\rm acc}\approx 2\times 10^{9}\,n_6^{-1}m_{10}^{-18/7}
M_{50}^{-5/7}v_{\rm ms,6}^{11/7}\,{\rm yr}\; .
\end{equation}
Thus, for a dense cluster ($n_6\sim 1$), or an especially high-mass
black hole ($M_{50}\gta 5$), significant growth can occur
in a Hubble time even in the absence of binaries in the core.  
Note that the growth rate accelerates with increasing
$M$, so once growth is significant the massive black hole will consume
the remaining field black holes quickly.
The ratio of three-body to two-body encounter rates is
\begin{equation}
{\nu_{3b}\over{\nu_{2b}}}\approx 2\times 10^4 M_{50}^{-5/7}v_{ms,6}^{4/7}\; .
\label{ratio}
\end{equation}
We see that unless there are virtually no binaries in the core into which a
massive black hole can exchange, such a black hole will encounter binaries
far more often than it encounters single stars.  However, note that merger
via three-body encounters may require tens to hundreds of interactions (see
below), whereas merger via direct two-body capture is essentially
immediate.  Moreover, just as three-body rates are modified by gravitational
radiation, the direct capture rate may be modified by deflections induced by
three-body effects.  Numerical simulations are required to determine when
the two-body growth rate exceeds the three-body growth rate of a massive
black hole in a cluster, and how the two effects couple.

\subsection{Mass of central black holes in globular clusters}

If the escape velocity from the core is 50~km~s$^{-1}$ (Webbink 1985)
and the field black hole is ejected with a kinetic energy a factor
$\approx
0.2(m/M)$ of the binary binding energy, then field black holes will be
ejected when the binary orbital radius is about $10^{13}m_{10}$~cm,
independent of the mass of the larger black hole.
At a fractional hardening of $0.2(m/M)$ per interaction, the number
of interactions required to bring the radius down to $\approx 3\times
10^{11}$~cm is $\approx 
\ln\left[10^{13}m_{10}/(3\times 10^{11}M_{50}^{1/2}m_{10}^{1/4})\right]/
\ln[1+0.2m/M]\approx 25M_{50}m_{10}^{-1}\ln(30M_{50}^{-1/2}m_{10}^{3/4})
\approx 100\,M_{50}$.
Thereafter it merges quickly by gravitational radiation.  Thus for each black
hole accreted in this manner, $\approx 100\,M_{50}$ are ejected, so the
massive black hole can in principle consume $\sim 0.1-1$\% of the  smaller
black holes that interact with it, ejecting the rest.  Here again the
precise eccentricity behaviour has a large impact; if high eccentricities
are reached in the process of hardening, the binary merges when the
semimajor axis is larger, and hence ejects fewer field black holes. The
smaller black holes will also interact with and eject each other,  further
cutting down the mass supply.  Portegies Zwart \& McMillan (2000) find that
interactions among the smaller black holes eject roughly half of them
within $\sim 2$Gyr, and about 90\% after several billion years.  Thus, if a
large black hole grows rapidly, it may be able to accrete a few tenths of a
percent of the initial mass in black holes.  How much mass would this be?

If the initial mass function of stars in the cluster is the
Salpeter IMF, $dN/dM\propto M^{-2.35}$ above $1\,M_\odot$ (and
flatter below one solar mass; see, e.g., Meyer et al. 2000), 
then the mass in stars above some threshold $M_0>1\,M_\odot$ is
\begin{equation}
\int_{M_0}^\infty M{dN\over{dM}}\,dM\propto M_0^{-0.35}\; .
\end{equation}
If black holes form from stars with initial masses $m>30\,M_\odot$, this
implies that $\sim$20\% of the initial mass might have been in stars that
form black holes.  Wind losses on the main sequence, mass loss during the
supernova, and ejection of some black holes by supernova recoil reduce
this fraction.  However, wind losses may be minimal for these
low-metallicity stars (Vink, de Koter, \& Lamers 2001), and supernova
recoil of black holes is expected to be small compared to that for neutron
stars (e.g., Sigurdsson \& Hernquist 1993), so mass losses could be
unimportant.  This is especially true  if stars of initial mass
$M>40\,M_\odot$ produce failed supernova (Fryer 1999) and hence lose
little mass.  Given the initial total mass in black holes and the fraction
lost to ejection, a large central black hole could accrete a fraction
$\sim 10^{-3}$ of the initial mass of the cluster.  Perhaps
coincidentally, this is comparable to the $\sim 0.5$\% ratio of
supermassive black hole mass to bulge mass found for many galaxies (e.g.,
Merritt \& Ferrarese 2001). Then, an initial cluster mass of $\sim
10^6\,M_\odot$, appropriate for the densest clusters, implies that the
black hole could gain a few hundred to a thousand  solar masses from just
the smaller black holes.  If there are multiple $\sim 50\,M_\odot$ black
holes that can merge with each other, or if high eccentricities are
reached, this number may become $\sim 10^3-10^4\,M_\odot$.

Once the small black holes are depleted significantly the main source of
fuel becomes the next most massive stars in the core. If the black holes
are consumed before $\gta 1.5\,M_\odot$ main sequence stars evolve, these
will dominate the core.  Otherwise, the next in line are neutron stars.
These neutron stars have a number density comparable to or smaller than
that of black holes (having a larger scale height  and not being
overwhelmingly more numerous) and are also less massive, so the
characteristic time for mass gain becomes longer by a large factor,
perhaps $\sim 100$ or more.  Thus, some extra growth is possible, but not
factors of many.  It is therefore plausible that this mechanism would
preferentially generate black holes with masses $M\sim 10^3\,M_\odot$.

\subsection{Detectability of black holes in clusters.}

Most globular clusters obviously do not have $10^{41}$~erg~s$^{-1}$ X-ray
sources.  If a substantial fraction contain high-mass black holes, why are
they not emitting profusely? To answer this, consider what happens when
stars of various types are accreted.  Also, remember that the most massive
type of star present in abundance will dominate the encounters. This is
because binary exchange interactions tend to yield a binary consisting of
the two most massive of the three stars involved, and a large number of
such interactions are required to harden the binary sufficiently to produce
a merger.  If, for example, 100 binary interactions are required before
merger, then the encounter rate with black holes must be $\lta 1$\% of the
total for the final binary to not include two black holes. Given that the
lower velocities of black holes increase their interaction cross section,
this would require severe depletion of the black holes in the core.  If
instead direct capture dominates, the higher mass and lower velocities of
massive objects give them a far higher cross section than the less massive
objects.  Thus, mergers are primarily with black holes when they are
present, then neutron stars, and finally white dwarfs and the remaining
main sequence stars (which have $M<0.8\,M_\odot$ for a typical current
cluster age).

A small black hole merging with a massive black hole will generate only
gravitational radiation.  The same is true with a neutron star.  The
critical orbital separation $r_{\rm tide}$ between an object of mass $M$
and one with mass $m\ll M$ and radius $R$, such that the less massive
object is destroyed by tidal forces, is $r_{\rm tide}\approx
2.5(M/m)^{1/3}R$, or about  $10^7$~cm for $M=50\,M_\odot$ and
$m=1.5\,M_\odot$.  Assuming that the massive black hole is not rotating
significantly (plausible, since interactions  are likely to be uniformly
distributed in direction),  the radius of the innermost stable circular
orbit, inside of which an object will plunge without further loss of
angular momentum, is $R_{\rm ISCO}=6GM/c^2=5\times 10^7\,M_{50}$~cm.   The
two radii are equal, $r_{\rm tide}=R_{\rm ISCO}$, when $M\approx
5\,M_\odot$, and since $R_{\rm ISCO}/r_{\rm tide}\propto M^{2/3}$ more
massive black holes swallow neutron stars even more easily.  Therefore, a
neutron star is not disrupted before merger.

A white dwarf is disrupted, but the merger is very quick.  A white dwarf
with typical mass $0.6\,M_\odot$ has a radius $R\approx 10^9$~cm (e.g.,
Panei, Althaus, \& Benvenuto 2000).   The tidal disruption radius is then
$\approx 10^{10}M_{50}^{1/3}$~cm, well outside the innermost stable
circular orbit.  The dwarf therefore transfers mass as in a low-mass
X-ray binary.  However, at such separations the time to merge via
gravitational radiation is short: roughly ten years for $M_{50}=1$,
scaling as $M^{-2/3}$.  Even if this time increases as the white
dwarf loses mass, there is only a short time in
which radiation can be detected.  In addition, the initial mass transfer
rate of $\sim 0.1\,M_\odot$~yr$^{-1}$ is $10^5$ times the Eddington rate
for a $50\,M_\odot$ black hole.  At such high rates emission may be in
the hypercritical regime in which neutrino emission dominates over photon
emission (Houck \& Chevalier 1991; Brown, Lee, \& Bethe 2000).  Thus,
white dwarfs, if accreted, will have a tiny duty cycle for observable
emission, if indeed the emission is observable at all.

A main sequence star of mass $\sim 0.6\,M_\odot$ has a radius 
$\sim$40 times larger than a white dwarf of the same mass.
The separation at disruption is thus 40 times higher, and the
gravitational merger time $(40)^4\approx 10^6$ times greater, than
for a white dwarf.  Thus, accretion of a main sequence star by a
massive black hole will produce copious photon emission.
At the Eddington rate a $0.6\,M_\odot$ star will be consumed by a
$\sim 50\,M_\odot$ black hole in $\sim 2\times
10^6$~yr, smaller than but comparable to the hardening timescale.
If a cluster black hole has exhausted its supply of compact objects
with mass greater than the main-sequence turn-off mass, it is therefore
possible that it may form an LMXB with a main sequence star.

Accretion from gas in the cluster is unlikely to be detectable.  
Although much of the primordial gas in globular
clusters may be swept out by supernovae, the winds from 
evolving red giants may be accreted.  If we assume that at any given
time 1\% of the stars are on the red giant branch, even a rich cluster
with $10^6$ stars will have only $10^4$ red giants, which are distributed
within a typical half-mass radius of 10~pc.  The mass loss rates for
low-mass red giants are difficult to determine but seem to be
$<10^{-8}\,M_\odot$~yr$^{-1}$ and may be significantly less for the
metal-deficient stars in globular clusters (Dupree 1986).
The winds come out in a large range of velocities broadly distributed
around the escape
velocity (Lamers \& Cassinelli 1999), which is roughly 50~km~s$^{-1}$ at 
the 1~AU radius of
a $0.8\,M_\odot$ red giant.  The winds take $3\times 10^{19}$
cm/$5\times 10^6$~cm~s$^{-1}$=$2\times 10^5$~yr to cross the volume,
so that on average each star donates $\approx 10^{-3}\,M_\odot$ in
wind material, for a total of $10\,M_\odot$ in ${4\pi\over 3}(10
\,{\rm pc})^3\approx 10^{59}$~cm$^3$.  The density is then
$2\times 10^{-25}$~g~cm$^{-3}$, consistent with the number density of
$\sim$0.1~cm$^{-3}$ inferred from pulsar dispersion measures in 47 Tuc
by Freire et al. (2001).  Suppose the central black hole captures
the gas via Bondi-Hoyle accretion, at the rate
\begin{equation}
{\dot M}\approx 2\times 10^{14}M_{50}^2
\rho_{-24} \left[v_6^2+c_6^2\right]^{-3/2}
\,{\rm g\ s}^{-1}\; ,
\end{equation}
where the hole has a velocity $10^6v_6$~cm~s$^{-1}$
relative to the cloud, and at infinity the density and sound speed
of the gas are $10^{-24}\rho_{-24}$~g~cm$^{-3}$ and $10^6c_6$~cm~s$^{-1}$,
respectively.  Then,
the accretion rate is $\approx 5\times 10^{13}$~g~s$^{-1}$ for a
$10^3\,M_\odot$ black hole.  This is, however, likely to be an overestimate
by several orders of magnitude.  The X-ray luminosity generated by accretion
ionises and heats its surroundings (e.g., Maloney, Hollenbach, \&
Tielens 1996), so that even well beyond the Bondi-Hoyle capture radius
the effective temperature is raised to $T>10^6$~K by Compton heating.
Pressure balance dictates that this heated gas must be less dense than
its surroundings by a factor proportional to the temperature, so that
the net accretion rate can be far lower than that indicated by a simple
Bondi-Hoyle estimate.  Accretion from any form of interstellar gas,
molecular, atomic, or otherwise, is thus unlikely to produce the
observed luminosity.  It is therefore difficult to rule out the presence
of a large black hole based on upper limits to X-ray emission.  For example,
Grindlay et al. (2001) have recently observed the cluster 47 Tuc with
the Chandra X-ray Observatory.  Even without including the possible
quenching of accretion by preheating, they find that despite the lack of any
detectable point source of X-rays within 3" of the cluster center,
a black hole with a mass of up to $500\,M_\odot$ could have escaped
detection.

A final way in which such a black hole could be detected in a globular
would be through its long-range effect on other cluster stars.  Current
data are ambiguous.  For example, Sosin (1997) finds that in the
post core collapse clusters M15 and M30 the radial density cusp of stars
is more consistent with the presence of a $\sim 10^3\,M_\odot$ black hole
than with simple equipartition core-collapse models.  However, there is
only marginal consistency with the expected cusp in the velocity dispersion
(e.g, Gebhardt et al. 1995).  Observations of the pulsars in M15 suggest
that there is no more than $\sim 700\,M_\odot$ in black holes in the
center of this core-collapsed cluster (Phinney 1993).  Future observations
will be required to place more stringent limits on the total black hole
mass in various clusters.

\section{Capture of Clusters by Their Host Galaxies and Accretion 
in the Galactic Disk.}

Globular clusters around a galaxy such as our own interact with the galaxy
through tidal stripping and other events.  Current estimates are that
approximately half of the clusters eventually merge with the disk and
eventually disperse (e.g., Gnedin \& Ostriker 1997; Gnedin, Lee, \& Ostriker
1999; Takahashi \& Portegies Zwart 1998, 2000).  
A massive black hole generated in the center of such a cluster is
then released and can interact with the interstellar medium. An intriguing
possible effect of a massive black hole inside a molecular cloud is that it
may help precipitate cloud collapse and star formation. Note that in a giant
molecular cloud of density $\rho=10^{-22}$~g~cm$^{-3}$, a $10^3\,M_\odot$
black hole contributes most of the mass out to a distance of a few parsecs.
The natural gravitational instability of the cloud would then be enhanced by
the presence of the hole, and in a few free-fall times $t_{\rm ff}\sim
1/\sqrt{G\rho}\approx 10^7(\rho/10^{-22}{\rm g\ cm}^{-3})^{-1/2}$~yr a star
cluster could form.  This is comparable to the crossing time of a giant
molecular cloud by a massive black hole, assuming a relative velocity of a
few km~s$^{-1}$ (appropriate if the black hole is nearly at rest relative to
the local average, since giant molecular clouds tend to have a random bulk
velocity of 3-5 km~s$^{-1}$; see Dame, Hartmann, \& Thaddeus 2001).  As in a
globular cluster, the most massive stars would sink to the center, where (as
main-sequence stars) they would accrete onto the black hole and generate
significant luminosity, in principle up to the Eddington luminosity.  This
may account for the observation by Matsushita et al. (2000) that the
brightest point X-ray source in M82 appears to be in a star cluster 
within a superbubble.

Note, however, that a black hole with a roughly stellar mass $M\approx
10\,M_\odot$ cannot easily grow to the required $\sim 10^3\,M_\odot$
masses in such an environment.  This is because the e-folding time for
mass increase at
even the Eddington rate is $6\times 10^7$~yr, comparable to or longer
than the timescale on which young star clusters dissipate (Portegies
Zwart et al. 2001), so that although a black hole may be luminous as
it feeds off of massive stars, the amount of mass it can accumulate is
relatively small.  These black holes must grow elsewhere.

How likely is it that a $\sim 10^3\,M_\odot$ black hole is in a dense
interstellar cloud at any given time, and what is the resulting
prediction for the number of active sources at any given time?
Consider first our Galaxy.  It currently has $\sim 150$ globular
clusters, of which $\sim$40\%, or 60, have core densities $n_6>0.1$
(Pryor \& Meylan 1993) and therefore produced massive black holes in
their cores according to our model.   We assume that a similar number
have merged with the Galaxy in its lifetime.  Molecular clouds with
densities $>10^2$~cm$^{-3}$ occupy a volume fraction $<10^{-2}$. It is
therefore not surprising that no massive black holes are  currently in
dense clouds in our Galaxy.

In contrast, galaxies undergoing active star formation are thought to
have surface densities of molecular  gas that are higher than ours by
a significant factor, perhaps $\sim 10$ or so (Taniguchi \& Ohyama
1998; Gao et al. 2001).  If the molecular clouds in star-forming
galaxies are similar to those in our Galaxy, the volume fraction of
dense clouds is therefore likely to be up by a similar factor, to
$\sim 10^{-1}$. That suggests that such a galaxy will have a few
massive black holes currently in star clusters within molecular
clouds, emitting $>10^{39}$~erg~s$^{-1}$.  This is consistent with the
number observed.

\section{Detection of gravitational waves from a 
$10^3\,M_\odot$ black hole in a globular cluster}

A major prediction of our model is that tens of percent of globulars
have $\sim 10^3\,M_\odot$ black holes.  As discussed in \S~2.4 and \S~3 these
holes are usually electromagnetically quiet.  However,
a massive black hole in a globular cluster around our Galaxy
is a potentially strong source of persistent gravitational waves in the
$10^{-4}-10^{-2}$~Hz sensitivity range of planned space-based
interferometers such as LISA.  In the final stages of inspiral
some of these sources may be detectable with ground-based detectors
such as LIGO~II up to $\sim 1$~Gpc away.
The strength, detectability, and number of such sources
depends on many uncertain parameters, such as the eccentricity
distribution and the current number density of lighter black holes
in globulars.  In this section we make rough estimates of the
type and strength of signals that may be found.

The dimensionless gravitational wave strain amplitude measured a distance
$r$ from a circular binary of masses $M$ and $m$ with a binary orbital
frequency $f_{\rm bin}$ is (Schutz 1997)
\begin{equation}
h=2^{5/3}(4\pi)^{1/3}{G^{5/3}\over c^4}f_{\rm bin}^{2/3}Mm(M+m)^{-1/3}
{1\over r}\; .
\label{amplitude}
\end{equation}
Note that the frequency of the gravitational waves is twice the
binary frequency.
A black hole binary will emit gravitational waves at all separations,
but here we focus on the portion of the inspiral during which gravitational
radiation has a significant influence on the orbital evolution.  From
\S2, this may happen in roughly the last $10^7\,M_{50}^{-1/2}$~yr of a total of 
$5\times 10^7\,M_{50}^{-1/2}$~years, for a duty cycle of $\sim 20$\%.
For a inspiral time of $10^7\,M_{50}^{-1/2}$~yr and a central density of 
$10^6n_6$~pc$^{-3}$, the semimajor axis is
$a\approx 4\times 10^{11}\,{\rm cm}n_6^{-1/4}M_{50}^{3/8}m_{10}^{1/4}$, 
so that the binary
orbital frequency is $f_{\rm bin}=4\times 10^{-5}{\rm Hz}\,
n_6^{3/8}M_{50}^{-1/16}m_{10}^{-3/8}$.  Numerically,
the gravitational wave amplitude is therefore 
\begin{equation}
h\approx 2\times 10^{-21}n_6^{1/4}M_{50}^{5/8}m_{10}^{3/4}(10\,{\rm kpc}/r)\; .
\end{equation}
The sensitivity of LISA  (from the LISA home page,
http://lisa.jpl.nasa.gov/science/science.html) in the range
$10^{-4}-10^{-3}$~Hz at a gravitational wave frequency 
$f_{\rm GW}=2f_{\rm bin}$
is approximately $h_{\rm min}=3\times 10^{-21}(f_{\rm GW}/10^{-4}\,{\rm
Hz})^{-5/2}=5\times 10^{-22}(f_{\rm bin}/10^{-4}\,{\rm Hz})^{-5/2}$
for a signal to noise ratio of 5 in an integration time of one year.  
Thus, the signal to noise ratio for most of the inspiral after the
last three-body interaction is
\begin{equation}
S/N=5\,h/h_{\rm min}\approx
5n_6^{19/16}M_{50}^{15/32}m_{10}^{-3/16}{10\,{\rm kpc}\over r}\; .
\end{equation}
Therefore, black holes with masses greater than $\sim 100\,M_\odot$ can be
detected with high significance in a year in a dense cluster within
10~kpc, and in a few years black holes with masses greater than $\sim
10^3\,M_\odot$ can be detected if $n_6>0.1$.  The steeply improving
sensitivity of LISA in the $10^{-4}-10^{-3}$~Hz frequency range means that
gravitational waves from a neutron star or white dwarf being accreted are
slightly easier to detect than black holes being accreted.  If 20\% of
mergers are in a favourable phase of the evolution, and if $\sim$40\% of
Galactic clusters have central number densities $n_6>0.1$ (Pryor \& Meylan
1993), then $\sim 10$ cluster black holes will be in this phase and
detectable if $M\gta 10^3\,M_\odot$.

The above applies to the amplitude of waves emitted during most of the
spindown.  In addition, some fraction of sources will be in the final
stage of their spindown, and hence will emit waves of higher amplitude.
For a given pair of masses the amplitude scales as $f_{\rm bin}^{2/3}$
and the sensitivity threshold of LISA scales as $f_{\rm bin}^{-5/2}$,
so the signal to noise scales as $f_{\rm bin}^{19/6}$ up to the 
$\approx 10^{-3}$~Hz sensitivity peak of LISA.  The fraction of
sources with a given orbital frequency scales as the inspiral time, which
is $\sim a^4\sim f_{\rm bin}^{-8/3}$.  Intermediate mass black hole binaries 
may therefore be detectable at large distances; for example, at the 15~Mpc
distance of the Virgo cluster a fraction $\sim 10^{-3}$ of such binaries
are likely to be observed, for a total of perhaps tens of
sources among the $\approx 10^3$ galaxies in the cluster.
Binaries in the last stage of inspiral will of course be visible to
even larger distances.  For example, from equation~(\ref{inspiral}) a
$10\,M_\odot$ black hole in a circular orbit around a $10^3\,M_\odot$
black hole at a binary orbital frequency of $5\times 10^{-3}$~Hz will
spiral in within $\approx 10$~yr.  The dimensionless strain amplitude
at a distance of 10~kpc would be $h=4\times 10^{-19}$, compared to the
LISA $S/N=5$ sensitivity of $h=10^{-23}$ at $10^{-2}$~Hz.  Such a signal
would therefore yield a 1 year signal to noise of 10 out to 200~Mpc with LISA.
Portegies Zwart \& McMillan (2000) estimate the number density of globulars
in the universe as $8.4h^3$~Mpc$^{-3}$; if $\approx 10$\% of these have
$10^3\,M_\odot$ black holes and core number densities $n_6\approx 1$,
we would expect $\approx 10$ such binaries to be detectable at any given
time.  

The merger itself is at frequencies above the range of LISA.  However, if
an intermediate mass black hole in a binary within a few years of merger is
found with LISA, the final inspiral could be anticipated and detected with
ground-based detectors such as LIGO, because the frequency, phase, and
location of the source would be known in advance. Unfortunately, the
frequency of this final merger [$f_{\rm GW}\approx
40\,(M/10^2\,M_\odot)^{-1}$~Hz at the last stable orbit of a nonrotating
black hole] is only marginally observable with LIGO-I, which has a lower
frequency cutoff $\sim 40$~Hz (see, e.g., Brady \& Creighton 2000).
However, LIGO-II is expected to be able to detect 10~Hz sources at 99\%
confidence for a 10 second integration and a dimensionless strain amplitude
of $h=2\times 10^{-23}$ (scaling from  Figure~3 of Brady \& Creighton
2000).  If black holes at the low end of the masses we  consider
($100\,M_\odot$ instead of $10^3\,M_\odot$) are present in a significant
number of globulars, their final merger is in the LIGO-II sensitivity
band.  From equation (\ref{inspiral}), a $10\,M_\odot$ black hole spiraling
into a $100\,M_\odot$ black hole spends approximately 10 seconds at a
gravitational wave frequency of 10~Hz or higher.  Using
equation~(\ref{amplitude}) as a rough guide, this signal would potentially
be detectable out to a gigaparsec or more.  Thus, advanced ground-based
detectors may pick up gravitational waves from a large number of $\sim
100\,M_\odot$ black hole mergers in globular clusters.

\section{Discussion and Conclusions}

We suggest that a population of $\sim 10^3\,M_\odot$ black holes is
generated in globular clusters and then released into the disks of their
host galaxies when the clusters are assimilated.  Subsequent 
accretion from young star
clusters formed from molecular clouds then produces the luminosity
observed from non-nuclear point sources in other galaxies. Such black
holes, accreting at less than the Eddington limit, can produce the
observed flux without strong beaming.

King et al. (2001) have proposed instead that the black holes are of
ordinary stellar mass but are strongly beamed.  They argue that larger
black holes would have difficulty in producing the required high-energy
X-rays if the emission is from blackbody annuli in the disk, because the
characteristic blackbody temperature for an Eddington accretor near the
inner edge of the disk is $kT\approx 2(M/M_\odot)^{-1/4}$~keV. However, we
point out that if a hot Compton corona exists near the black hole, as is
usually invoked to explain the highest energy emission, the temperature
scales as $M/r$ and is thus independent of mass, so unless the spectrum is
clearly of the multicolour blackbody form and has no nonthermal tail the
X-ray spectrum does not easily distinguish between beamed and unbeamed
emission.  For example, we note that the Galactic microquasars
GRS~1915+105 and GRO~J1655-40 have significantly harder emission than
would be expected from a simple multicolour blackbody (Makishima et al.
2000).  We also point out that blazars, which are believed to have
emission beamed towards us, tend to have relatively flat $\nu F_\nu$
spectra, varying typically by only a factor $\sim 10$ from
$10^{10}-10^{18}$~Hz (up to an X-ray to radio ratio of $10^3$ for some
Einstein Slew Survey blazars; see Fossati et al. 1998).  In contrast, the
X-ray to radio $\nu F_\nu$ ratio for Galactic microquasars, which are not
beamed towards us, is much greater; $10^6$ for GRS~1915+105 (Ogley et al.
2000 for radio; Rao et al. 2000 for X-ray) and $10^5$ for GRO J1655-40
(Hannikainen et al. 2000 for radio; Zhang et al. 1997 for X-ray). These
latter sources are comparable to the brightest X-ray source in M82, which
has an X-ray to radio ratio of at least $10^5$ (see Matsumoto et al. 2001
and Kronberg et al. 2000 for radio; Kaaret et al. 2001 for X-ray).
Although this does not prove that the intermediate mass black hole
candidates are unbeamed, it does argue in that direction.

Additional observations would be very helpful in determining whether
beamed or unbeamed models are favoured.  Looking at radio spectra of these
sources, at high angular resolution, will allow a broader sample to be
compared with blazar spectra.  Another difference is the expected minimum
time scale of variability.  For an unbeamed black hole of, e.g.,
$10^3\,M_\odot$, the minimum time scale would be the light crossing time
across the diameter of the minimum stable orbit, or $\approx 0.1$~s. The
time scale is likely to actually be a few times this; for example,
Cyg~X-1, with a black hole mass likely to be $\sim 10\,M_\odot$, has a
minimum timescale of significant variability of $\approx 3$~ms, three
times the light crossing time (Revnivtsev et al. 2000).  Above this
timescale, the fractional variability increases significantly. For a
beamed black hole of mass $\sim 10\,M_\odot$, the minimum time scale is
decreased by a factor of $\sim 10^3$ by relativistic effects and the
smaller size of the hole itself.  A long observation with Chandra in
continuous clocking mode, with a time resolution of 3~ms, would help
distinguish between beamed and unbeamed models.  Such observations are
important to determine whether the bright sources are the first
representatives of a third class of black holes.

\section*{Acknowledgements}

We thank Sterl Phinney, Derek Richardson, and especially Steinn
Sigurdsson for discussions.  We also thank Andrew Wilson and Andy Young
for discussions and comments on a previous version of this manuscript.
This work was supported in part by NASA grant NAG 5-9756.

\end{document}